\documentclass{tex-library/jfm}

\usepackage{xcolor}
\usepackage{amsfonts,amssymb,amsmath,amsthm}
\usepackage[colorlinks]{hyperref}
\usepackage[capitalise]{cleveref}
\usepackage{graphicx}

\usepackage[official]{eurosym}
\usepackage{mathtools}

\usepackage{lipsum}

\usepackage[english]{babel}

\newcommand{\tR}[1]{#1}

\usepackage{mathptmx}

\DeclareMathAlphabet{\itbf}{OML}{cmm}{b}{it}
\DeclareMathAlphabet\mathbfcal{OMS}{cmsy}{b}{n}

\newcommand{\RB}{Rayleigh-B\'enard}
\newcommand{\RBC}{\RB{} convection}
\def\RaN{\mbox{Ra}}
\def\ReN{\mbox{Re}}
\def\PrN{\mbox{Pr}}
\def\NuN{\mbox{Nu}}

\def\bU{{\itbf U}}

\def\bu{{\itbf u}}

\shorttitle{Rayleigh-B\'enard thermal convection perturbed by a horizontal heat flux}
\shortauthor{J. M. Huang and J. Zhang}

\title{Rayleigh-B\'enard thermal convection perturbed by a horizontal heat flux}

\author{Jinzi Mac Huang\aff{1,2}\corresp{\email{machuang@nyu.edu}} and Jun Zhang\aff{1,2,3}\corresp{\email{jz11@nyu.edu}}}

\affiliation{
\aff{1}NYU-ECNU Institute of Physics and Institute of Mathematical Sciences, New York University Shanghai, Shanghai, China
\aff{2}Applied Math Lab, Courant Institute, New York University, New York, USA
\aff{3}Department of Physics, New York University, New York, USA
}

\begin{document}
\maketitle

\begin{abstract}
In \RBC{}, it has been found that the amount of heat passing through the fluid has a power law dependence on the imposed temperature difference. Modifying this dependence, either enhancing or reducing the heat transfer capability of fluids, is important in many scientific and practical applications. Here, we present a simple means to control the vertical heat transfer in \RBC{} by injecting heat through one lateral side of the fluid domain and extracting the same amount of heat from the opposite side. This horizontal heat flux regulates the large-scale circulation, and increases the heat transfer rate in the vertical direction. Our numerical and theoretical studies demonstrate how a classical \RBC{} responds to such a perturbation, when the system is near or well above the onset of convection. 
\end{abstract}

\begin{keywords}
Authors should not enter keywords on the manuscript, as these must be chosen by the author during the online submission process and will then be added during the typesetting process (see http://journals.cambridge.org/data/\linebreak[3]relatedlink/jfm-\linebreak[3]keywords.pdf for the full list)
\end{keywords}

\section{Introduction}

There are many similarities between thermal, electronic, and fluid systems. For example, Ohm's law relates a flux of charges to an electrostatic potential difference, while Fourier's law of heat conduction $|q| \propto \Delta T$ states that heat flux $|q|$ is proportional to the temperature difference $\Delta T$. Other physical laws also share similar mathematical forms, such as the conservation laws \citep{Evans2010} regarding electrical charge, energy, and mass. Drawing analogies between these different systems has revealed many mechanisms that are known in one system but less obvious in the other \citep{schonfeld1954analogy}. Examples include the fluid ``transistor'' circuits proposed more than 60 years ago \citep{Pursglove1960} and a recent example on the flow rectifier based on Nikola Tesla's concept of a fluidic diode \citep{nguyen2021early,nguyen2021tesla}.

\begin{figure}
  \centerline{\includegraphics[width=\textwidth]{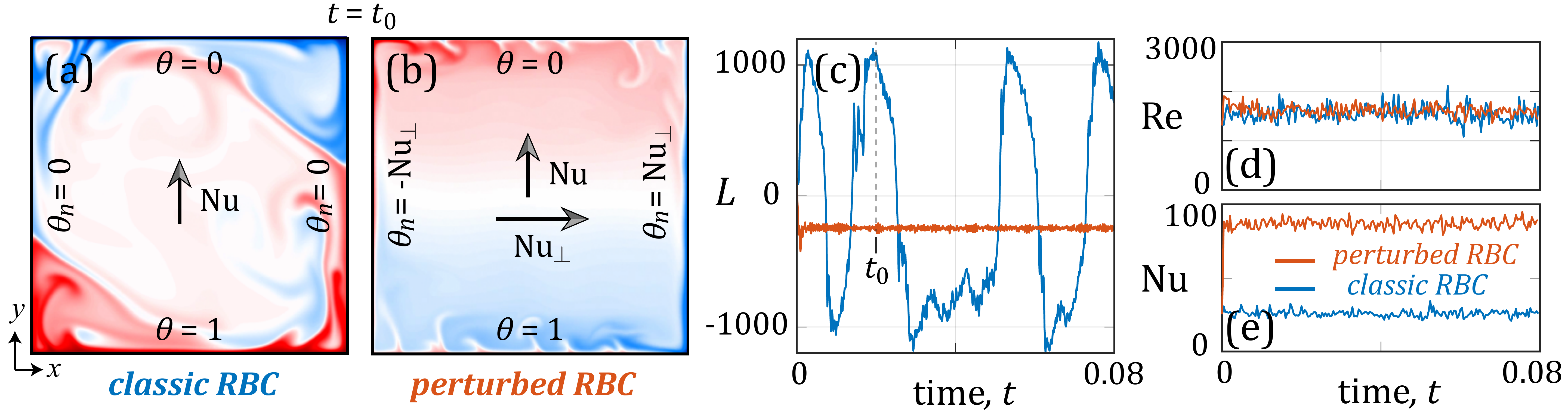}}
  \caption{At $\RaN = 10^8$ and $\PrN = 4.4$, flow and thermal structures of (a) A classic RBC; (b) A perturbed RBC, where a horizontal heat flux is added through sidewall flux conditions. \tR{The dimensionless temperature is $\theta$, and its wall normal derivative is $\theta_n$.} (c) In classic RBC, the total angular momentum $L(t)$ (blue data) alternates between positive and negative values due to the reversals of the large-scale circulation. In the perturbed case (red data), $L(t)$ stays negative and stable as the large-scale circulation is always clockwise (CW). (d) Defined by the maximum flow velocity (\cref{sec2}), the Reynolds number of both the classic and perturbed RBC is similar. (e) The Nusselt number is significantly enhanced by introducing the horizontal flux. The horizontal flux in (b)-(e) is $\NuN_\perp = 128$. The time $t=t_0$ of snapshots (a) and (b) is marked in (c), and full videos of (a) and (b) are included in the Supplements.}
\label{fig1}
\end{figure}

Combining these systems also reveals nontrivial and sometimes surprising dynamics. For example, allowing mass flow to a thermal system introduces nonlinearity that Fourier's law does not capture. It is well known that thermal convection appears when temperature gradient exceeds a threshold \citep{busse1978non,niemela2000turbulent,Childress2009}. One canonical example is the \RBC{} (RBC), whose configuration is simple: a closed domain of fluid is subject to heating from the bottom and cooling from the top, while the sidewalls remain adiabatic, as shown in \cref{fig1}(a). The cooled fluid near the top plate is heavier than the heated fluid near the bottom, creating an instability under gravity. Convection occurs when this system is beyond a threshold or onset, and the vertical heat flux has been observed to depend on the temperature difference nonlinearly: $|q| \propto \Delta T^\gamma$ where $\gamma$ is around 1.3, for a wide range of external and fluid parameters \citep{Belmonte1994,grossmann2000scaling,Ahlers2009}.

There have been numerous attempts to modify the heat transport in RBC and deviate from the above dependency, as a reduced or enhanced heat transfer rate is sometimes desired in applications like effective ventilation or energy preservation. Successful experimental examples include adding rotation to the RBC \citep{Stevens2009,Zhong2010a}, where moderate rotation rates enhance the heat flux and high rates reduces heat flux; changing the surface roughness of the top and bottom plates \citep{PhysRevLett.81.987}, where corrugated surface patterns result in higher heat transfer. Tilting the convection cell \citep{PhysRevFluids.3.113503,Guo2015} or inserting insulating partitions into the fluid domain \citep{Bao2015} also leads to regulated circulation and higher overall heat transfer rate. Combined experimental and numerical efforts have also brought insights on heat transport enhancement through coherent structure manipulation \citep{Chong2017}.

The examples above show the possibility of modifying the heat transfer in \RBC{}, and many of them involve adding mechanical parts or changing the geometry of the convection cell. Without moving parts, an active control of \RBC{} through modified boundary conditions is also possible. Examples include the work of \cite{howle1997active}, where the unstable convetive fluid motion is stabilized through an active heat flux imposed through the boundary conditions, and the study by \cite{zhang2020controlling}, where the reversal of large-scale circulation is suppressed by introducing small control regions on the sidewalls.

In order to externally control and enhance the heat transfer, we impose a horizontal heat flux $q_\perp$ to the classical \RBC{} as shown in \cref{fig1}(b), by setting sidewall temperature gradients as $|\partial_n T| = q_\perp/K$, where $K$ is the thermal conductivity. This additional flux, by heating on one side and cooling on the opposite side, modifies the bulk fluid motion by inducing buoyancy jets near the sidewalls, resulting in a unidirectional large-scale circulation as seen in \cref{fig1}(b), where the fluid angular momentum $L$ (defined in \cref{sec2}) ceases to alternate as shown in \cref{fig1}(c). Interestingly, the overall magnitude of flow angular momentum is reduced while the flow velocity, measured by the Reynolds number (\cref{sec2}), stays nearly unchanged in \cref{fig1}(c)-(d). On the other hand, the vertical heat transfer rate measured by the Nusselt number (\cref{sec2}) in \cref{fig1}(e) is greatly increased compared to the classic case.

To understand these observations, we examine the side-heated and side-cooled \RBC{} systematically through two-dimensional (2D) numerical simulations. Considering that the flow and thermal structures involved are largely 2D, especially when side heating and cooling dictate the flows, the heat transfer properties of \RBC{} can be accurately accounted for by such numerical simulations \citep{schmalzl2002influence,schmalzl2004validity,Ahlers2009}. \tR{In \cref{sec2}, the mathematical formulation and numerical implementation are introduced. The main results of our study are presented in \cref{sec3}, which includes a detailed investigation of how horizontal heat flux affects the flow and temperature fields (\cref{sec3.1}), how bulk quantities respond to this perturbation (\cref{sec3.2}), and how time dependent perturbation alters the dynamics of \RBC{} (\cref{sec3.3}). Finally, we will discuss some potential applications in \cref{sec4}, where a simple mechanism is used to actively control the net circulation of \RBC{}. }

\section{Numerical setup}
\label{sec2}
To formulate our problem dimensionlessly, we rescale temperature $T$ by $\Delta T = T_b-T_t$ so the dimensionless temperature is $\theta = (T-T_t) / \Delta T $, and rescale coordinates $(X,\,Y)$ by $L$ (domain height) so the coordinates $(x,\,y)\in \Omega \triangleq (0,1) \times (0,1)$, as shown in \cref{fig1}(a). The time is rescaled by the thermal diffusion time scale of $L^2/\kappa$ where $\kappa$ is the thermal diffusivity of the fluid, leading to the dimensionless velocity field $\bu = \bU L/\kappa$. Two relevant dimensionless numbers are the Rayleigh number $\RaN =  \alpha g \Delta T L^3 / \kappa \nu$, which measures the relative strength between the thermally induced buoyancy force and the fluid viscous force, and the Prandtl number $\PrN = \nu / \kappa $, the ratio between fluid viscosity and thermal diffusivity. Here $\rho, \nu, \alpha$ and $g$ are the density, kinematic viscosity, and thermal expansion coefficient of the fluid and the acceleration due to gravity, respectively.

The heat flux passing through the vertical direction can be non-dimensionalized as the Nusselt number, $\NuN = q / q_c = - \int_{0}^1\partial_y \theta (x,\,1) dx$, which is the ratio between the convective flux $q  = - (K/L) \int_{0}^L\partial_Y T (X,\,L) dX$ and the conductive flux $q_c = K \Delta T / L$. To simplify our notation, we also define a horizontal Nusselt number as $\NuN_\perp =  q_\perp L/K\Delta T $. The symbol $\perp$ indicates that the horizontal heat flux is perpendicular to the traditionally vertical heat flux ($\NuN{}$) in \RBC{}. \tR{As $\NuN_\perp$ and $\NuN$ are both rescaled by the conductive flux $q_c$, we can define $S = \NuN{}_\perp/\NuN{}$ and directly compare the horizontal and vertical heat fluxes. This makes $\NuN_\perp$ a more natural and direct choice of measuring the imposed horizontal flux, compared to other dimensionless numbers such as the horizontal flux Rayleigh or Grashof number. }

Another important number is the Reynolds number $\ReN{}=U_{\mbox{max}}L/\nu = \PrN^{-1} u_{\mbox{max}}$, where the maximum velocity $u_{\mbox{max}} = \max |\mathbf{u}|$ represents the flow speed scale. In order to investigate the strength of large-scale circulation, we also define the dimensionless total angular momentum as the integral $L(t) = \int_0^1\int_0^1 \left[(x-0.5)v(x,\,y,\,t)-(y-0.5)u(x,\,y,\,t)\right]\, dx\,dy$, so a positive $L$ represents counter-clockwise (CCW) large-scale circulation.

Overall, the system has three control parameters, or \textit{inputs}: $\RaN$, $\PrN$ and $\NuN_\perp$. \textit{Outputs} like $\NuN{}$, $\ReN{}$, and $L$ are functions of these inputs. To further simplify our study, we set $\PrN{} = 4.4$ (water at $40 ^\circ$C) for all the simulations, as varying $\PrN{}$ in the range of $1-10$ does not significantly change our results.

Denoting the vorticity as $\omega = \hat{\itbf{k}}\cdot \nabla \times \bu$ and the stream function as $\psi$ such that $\bu = \nabla_\perp \psi = (\psi_y,\,-\psi_x)$, we can write the Navier-Stokes equation in the vorticity-stream function format
\begin{align}
\label{temp-eqn}
\frac{\partial \theta}{\partial t} + \bu\cdot\nabla \theta &=  \Delta \theta \quad \mbox{in} \quad \Omega,\\ 
\frac{\partial\omega}{\partial t} + \bu\cdot\nabla \omega &= \PrN \Delta \omega   + \PrN \RaN \frac{\partial \theta}{\partial x} \quad \mbox{in} \quad \Omega,\label{sec-eqn} \\
 -\Delta \psi = \omega,\ \  \bu &= \nabla_\perp \psi \quad \mbox{in} \quad \Omega, \label{last-eqn}
\end{align}
with boundary conditions
\begin{equation}
\begin{cases}
& \psi = \psi_n = 0 \quad \mbox{on} \quad \partial \Omega,\\ 
& \theta = 0 \quad \mbox{on} \quad \partial \Omega_{\textup{up}}, 
\quad \theta = 1 \quad \mbox{on} \quad \partial \Omega_{\textup{down}},\\
& \dfrac{ \partial \theta}{\partial n} = -\NuN_\perp \  \mbox{on} \  \partial \Omega_{\textup{left}},
\quad \dfrac{ \partial \theta}{\partial n} = \NuN_\perp \  \mbox{on} \  \partial \Omega_{\textup{right}}.
\end{cases}
\end{equation}
\tR{Here $\mathbf{n}$ is the outward normal vector, so $\partial_n = -\partial_x$ on $\Omega_{\textup{left}}$ and $\partial_n = \partial_x$ on $\Omega_{\textup{right}}$.}

To solve these equations, we employ a pseudo-spectral scheme \citep{Peyret2002} that uses the Chebyshev method. Spatial variables are discretized on the Chebyshev nodes, with operations like derivatives and integrations performed through corresponding discrete operators. Moreover, an efficient anti-aliasing filter \citep{Hou2007} is applied when evaluating nonlinear advection terms pseudo-spectrally. At each timestep, a second-order implicit-explicit Adam-Bashforth Backward-Differentiation method solves for the stiff parabolic equations and the nonlinear equations. Typically, the simulation has $N=200$ Chebyshev nodes in each dimension, and the timestep is set as $\Delta t = 2\times10^{-4} \cdot\mbox{Ra}^{-1/2}$ [considering $|\bu|\sim$ Ra$^{1/2}$ \citep{Ahlers2009}] to maintain numerical accuracy and stability. \tR{These parameters are tested to yield spatially and temporally resolved solutions for \cref{temp-eqn}-\cref{last-eqn}. As the unevenly-spaced $N$ Chebyshev points has an $O(N^{-2})$ density near the boundary \citep{trefethen2000spectral}, the small-scale boundary layer structure can be efficiently resolved there, allowing us to simulate high $\RaN{}$ convection with a small number of Chebyshev nodes. }

\tR{At each $\RaN{}$, we first evolve the classic \RBC{} ($\NuN{}_\perp = 0$) to a dynamical equilibrium state, which is used as the initial condition for simulations of different $\NuN{}_\perp$. Each simulation then runs for 4 million time steps, and time averaged quantities such as $\NuN{}$ and $\ReN{}$ are averaged during the latter 2 million time steps, where the system has equilibrated.}

\section{Results}
\label{sec3}
\subsection{Flow and temperature profiles}
\label{sec3.1}
\begin{figure}
  \centerline{\includegraphics[width=\textwidth]{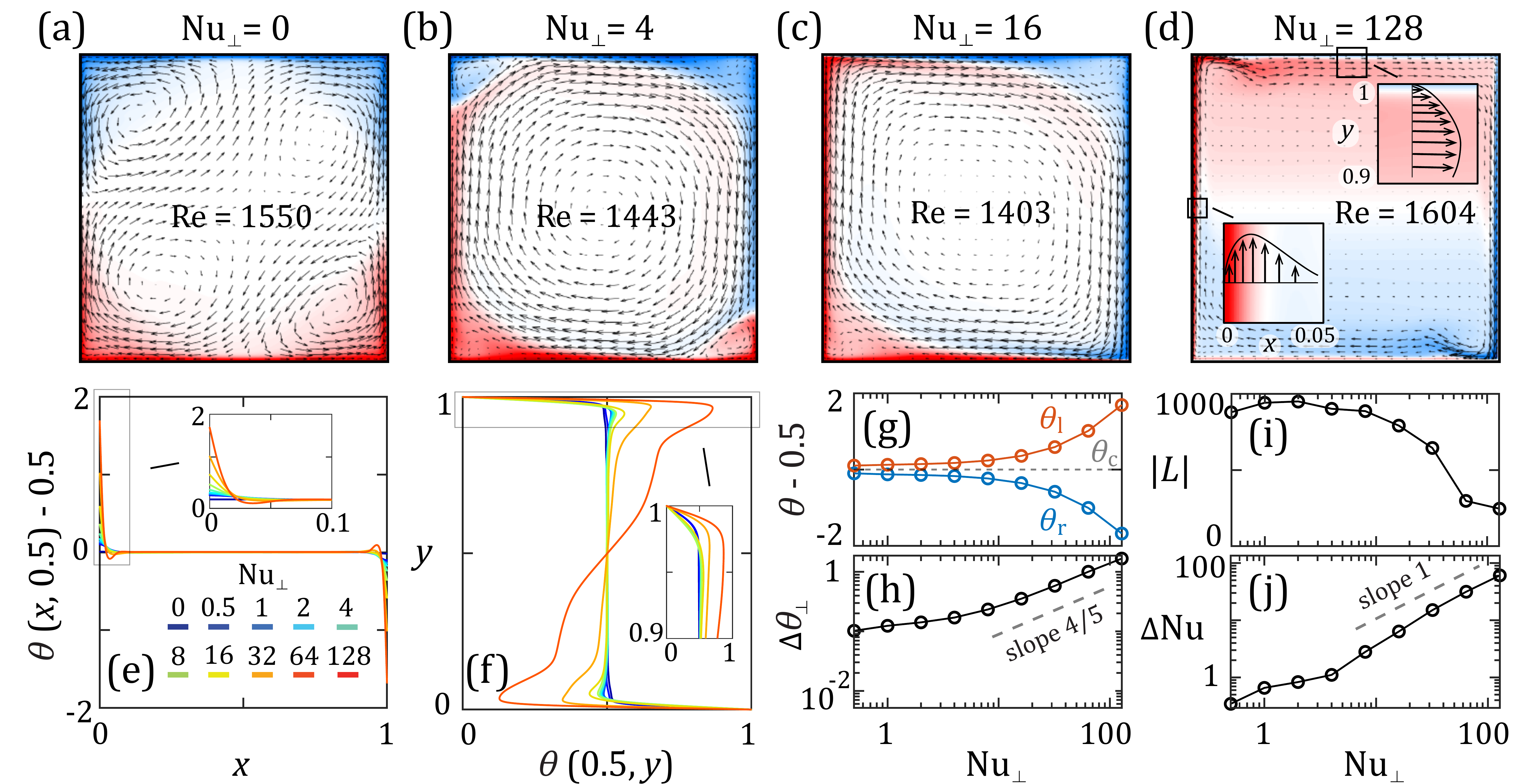}}
  \caption{ Flow (arrows) and temperature (color map) profiles of \RBC{} with an additional horizontal heat flux. (a)-(d) Time averaged flow and temperature fields at $\RaN = 10^8$ with four different strengths of horizontal flux. \textit{Insets} of (d) show the boundary layer structure for the flow and temperature fields near the top center $(0.5,\,1)$ and left center $(0,\,0.5)$. Bottom and right boundary layers are symmetric to the top and left boundary layers with respect to the center $(0.5,\,0.5)$. (e) Horizontal temperature profiles along $y=0.5$. (f) Vertical temperature profiles along $x = 0.5$.  (g) The time-averaged left wall temperature $\theta_l$ increases while the right wall temperature $\theta_r$ decreases symmetrically, about bulk $\theta_c$, with increasing $\NuN_\perp$. (h) At high $\NuN_\perp$, the horizontal temperature change  $\Delta \theta_\perp = \theta_l-\theta_c = \theta_c - \theta_r$ takes the $4/5$ power law as an asymptote. (i) Time-averaged total angular momentum $|L|$ decreases with the horizontal flux. (j) The Nusselt enhancement $\Delta\NuN{} = \NuN- \NuN_0$ scales linearly with the horizontal flux. }
\label{fig2}
\end{figure}

The effect of adding a horizontal heat flux to \RBC{} is directly reflected in the temperature and flow fields. \Cref{fig2}(a)-(d) show four time-averaged flow and temperature profiles at different magnitudes of horizontal flux. At a constant $\RaN = 10^8$, the Nusselt number for the classical \RBC{} ($\NuN_\perp=0$) is $\NuN_0\approx 25$, and the direction of circulation is not deterministic [\cref{fig1}(c) and \cref{fig2}(a)]. A clear influence of the horizontal flux is already present at $\NuN_\perp = 4$ [\cref{fig2}(b)], where the direction of circulation becomes deterministically clockwise due to the buoyancy driven flows near the vertical walls. For $\NuN_\perp \geq 16$ [\cref{fig2}(c)-(d)], the profiles of temperature and flow fields are dictated by the sidewall heating and cooling and the corner rolls are eliminated, similar to the profiles in a horizontal convection without bottom heating or top cooling \citep{Belmonte1995}. As a measure of relative strength between horizontal and vertical flux, $S = \NuN{}_\perp/\NuN{}$ is in the range of 0 - 1.5 in \cref{fig2}.

With increasing $\NuN_\perp$, the Reynolds number stays at a similar level despite the increasing sidewall heating, which only seems to thin the vertical boundary layers near the sidewalls as shown in \cref{fig2}(b)-(d). Another consequence of this thinning of boundary layers can be seen in \cref{fig2}(i), where the time-averaged total angular momentum $|L|$ decreases with increasing $\NuN{}_\perp$. This response seems to be counter-intuitive at first, but can be explained after examining the flow fields. As the magnitude of flow speed stays at a similar level, a thin boundary layer and a lack of bulk circulation [\textit{inset} of \cref{fig2}(d)] lead to a reduced bulk contribution to the total angular momentum $|L|$.

The time-averaged temperature distributions along the horizontal center line $y=0.5$ and the vertical center line $x=0.5$ are shown in \cref{fig2}(e)-(f). On the horizontal cut $y=0.5$, the temperature stays at $\theta(x,\,0.5) = 0.5$ when no horizontal flux is added. As the sidewall heating-cooling increases, the temperature on the heating wall becomes higher while the temperature on the cooling wall becomes symmetrically lower. This change of temperature generates buoyancy jets along the two sidewalls, which can also be seen in \cref{fig2}(a)-(d). Due to flow advection, thermal boundary layer develops near the sidewalls as shown in \cref{fig2}(d), similar to the boundary layer profile near a heated or cooled vertical wall in an infinite space \citep{Schlichting2016}. As the buoyancy jets move upward/downward, they are diverted to the right/left as they meet the top/bottom wall, resulting in an overall clockwise circulation. Due to the relatively hot/cold fluid coming from the region near the left/right wall, the vertical temperature profile $\theta(0.5,\,y)$ in \cref{fig2}(f) has an inversion with a temperature rise near the top wall and a drop near the bottom. Namely, the bulk fluid becomes increasingly stratified. As a result, the magnitude of the temperature gradient near the top and bottom wall is increased, as shown in the inset of \cref{fig2}(f), leading to a higher vertical heat transfer rate. \tR{Even though the same amount of heat flowing into the left sidewall leaves from the right, this horizontal heat flux does alter the overall flow structures and affect the boundary layers, causing the strong Nu enhancement shown in \cref{fig1}(e) without significantly changing the flow speed.}

Defining $\theta_l = \theta(0,\,0.5)$, $\theta_r = \theta(1,\,0.5)$, and $\theta_c = \theta(0.5,\,0.5)$ to be, respectively, the temperature of the left wall, right wall, and bulk center, their time-averaged values at different $\NuN_\perp$ are shown in \cref{fig2}(g). The temperature rise and drop of the left and right wall are symmetric with respect to the bulk center, which has a mean temperature $\theta_c = 0.5$. Defining $\Delta \theta_\perp = \theta_l-\theta_c = \theta_c-\theta_r$ to be the temperature change on the sidewall, it increases with the horizontal flux $\NuN_\perp$ as shown in \cref{fig2}(h). Asymptotically, the measured data approaches $\Delta\theta_\perp \propto \NuN_\perp^{4/5}$ when $\NuN_\perp$ is high. This is consistent with the boundary layer scaling between the temperature rise $\Delta T_\perp$ and the injected heat $Q$ from a vertical heated wall, $\Delta T_\perp \propto Q^{4/5}$ \citep{Schlichting2016}. \tR{In this limit, we also note that $\Delta\theta_\perp$ can be greater than 1 as shown in \cref{fig2}(g), where the horizontal temperature difference is greater than the vertical one and the horizontal flux becomes the main driver of thermal convection.}

Focusing on the enhancement of $\NuN$, \Cref{fig2}(j) shows the Nusselt increment $\Delta \NuN = \NuN - \NuN_0$ increases monotonically with  $\NuN_\perp$. Interestingly, the increment $\Delta\NuN$ scales linearly with $\NuN_\perp$. An explanation for this is that the high and low temperature jets generated by the left and right sidewalls must pass by the top and bottom plates, leading to higher temperature gradients in the boundary layers shown in \cref{fig2}(f). Overall, the horizontal flux gradually limits the fluid circulation to a thin boundary layer region near four boundaries as shown in \cref{fig2}(d), and the heat conduction within the boundary layer strengthens the heat transfer rate ($\NuN{}$) in the vertical direction. In the following section, we will systematically review such a Nusselt enhancement at different $\RaN$.

\subsection{Bulk heat transfer and flow properties}
\label{sec3.2}
\begin{figure}
  \centerline{\includegraphics[width=\textwidth]{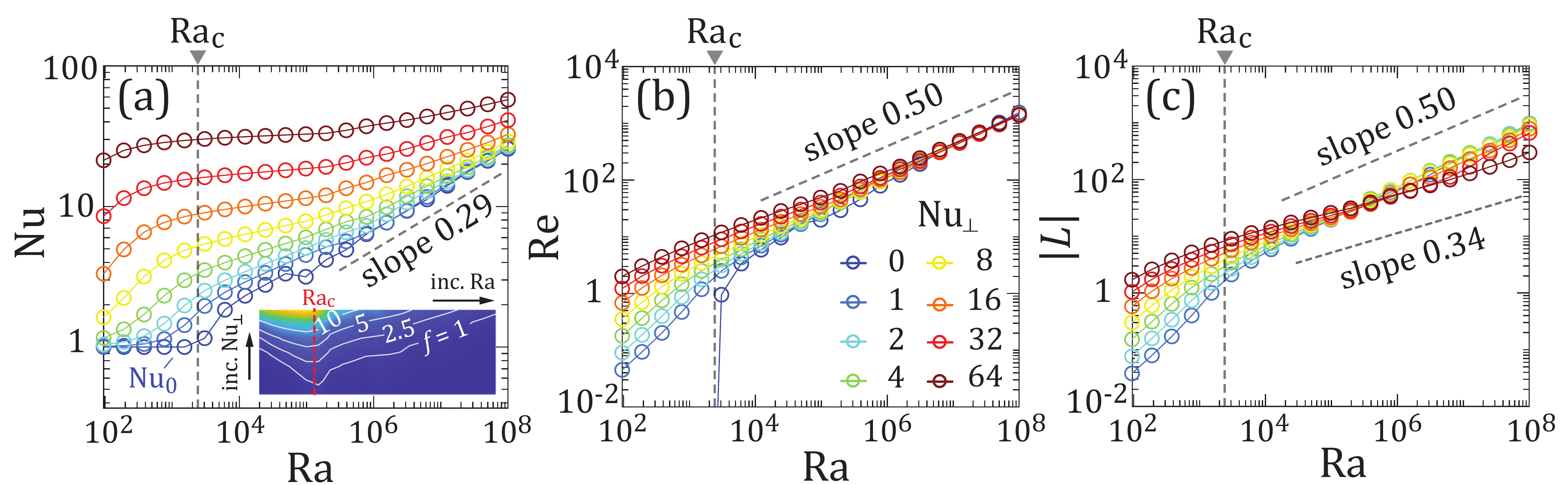}}
  \caption{Time-averaged bulk quantities, $\NuN{}, \ReN{}$, and $|L|$ measured at various Rayleigh number $\RaN$ and horizontal heat flux $\NuN_\perp$. (a) The Nusselt number increases with $\NuN_\perp$ at a given $\RaN$. The dark blue curve $\NuN_0$ corresponds to the classic \RBC{} with adiabatic sidewalls. \textit{Inset}: Relative $\NuN$ enhancement $f = (\NuN/\NuN_0)-1$ reaches a local maximum near $\RaN_{\mbox{c}} = 2415$. Arrows indicate the direction of increasing $\RaN$ and $\NuN{}_\perp$, where $\RaN\in[10^2,\, 10^8]$ and $\NuN{}_\perp\in[0,\, 64]$. (b) Horizontal flux leads to nonzero Reynolds number even for $\RaN<\RaN_{\mbox{c}}$, and $\ReN\sim\RaN^{0.5}$ at high $\RaN$ is consistent with the scaling in classic \RBC{} \citep{Ahlers2009}. (c) Total angular momentum $|L|$ has a 0.5 power law scaling with $\RaN{}$ when the horizontal flux is small, but takes a 0.34 power law when $\NuN_\perp$ dominates. }
\label{fig3}
\end{figure}

Without the horizontal flux ($\NuN_\perp = 0$), $\NuN_0$ is known to depend on the Rayleigh number in a nonlinear way: When the Rayleigh number is below a critical number [around $1708$ \citep{koschmieder1993benard}], the viscous force suppresses fluid motion and the fluid behaves like solid, hence $\NuN_0 = 1$; Increasing the Rayleigh number well beyond critical, a power law relation emerges as $\NuN_0\propto \RaN^{0.29}$ despite local deviations. The more detailed dependence between $\NuN$ and $\RaN$ is summarized in the theory of Grossmann \& Lohse \citep{grossmann2000scaling}. In our simulation, this relation is recovered as the dark blue data of \cref{fig3}(a). As we shall determine later, the critical Rayleigh number in our numerical system is near $\RaN_{\mbox{c}} = 2415$, slightly higher than the theoretical value 1708 for an infinitely wide convection cell \citep{koschmieder1993benard}, but consistent with experiments and simulations conducted in finite domain \citep{PhysRevE.81.046318}. 
 
The behavior near $\RaN_{\mbox{c}}$ changes with the addition of a horizontal flux: \Cref{fig3}(a) shows that the Nusselt number can be greater than 1 when $\RaN < \RaN_{\mbox{c}}$ and $\NuN_\perp > 0$. The side heating and cooling set up the circulation even though the vertical temperature gradient is small, and this mixing effectively stirs the bulk fluid and increases the vertical flux. Indeed, \cref{fig3}(b)-(c) show that horizontal flux does lead to fluid motion with non-vanishing $\ReN$ and $L$, even at $\RaN < \RaN_{\mbox{c}}$. Paying attention to the top half of the convection cell, the warm fluid close to the left wall flows upward to the top plate, creating a larger temperature difference between the fluid and the top plate and a smaller characteristic length as the boundary layer thickness decreases with the fluid velocity [\cref{fig2}(f)]. All together, these effects result in a greater temperature gradient near the top hence an enhanced $\NuN$. Similar analysis can be performed symmetrically near the bottom plate, as the system is symmetric about its center such that $\theta(x,\,y) = 1 - \theta(1-x,\,1-y)$.


This enhancement of $\NuN$ exists for a wide range of $\RaN$ and $\NuN_\perp$. Shown in the \textit{inset} of \cref{fig3}(a), the relative enhancement $f = (\NuN/\NuN_0)-1$ is an increasing function of $\NuN_\perp$ when $\RaN$ is fixed. At a fixed $\NuN_\perp$, however, the value of $f$ peaks around the critical Rayleigh number $\RaN_{\mbox{c}}$. One possible explanation for this peak is that the fluid (and hence the vertical heat flux) is easily perturbed when $\RaN\sim\RaN_{\mbox{c}}$ while the unperturbed $\NuN_0$ is still at unity. As $\RaN$ increases, the relative enhancement $f$ becomes smaller and approaches 0 asymptotically. This is understandable as the relative strength of the imposed horizontal flux and the unperturbed vertical flux $\NuN_\perp/\NuN_0$ diminishes with increasing $\RaN$. Eventually, $\NuN_0\gg\NuN_\perp$ as $\RaN\to\infty$ so the horizontal flux has negligible effects on the \RBC{}. In the same limit, the Reynolds number shown in \cref{fig3}(b) also returns to the asymptote of $\ReN{}\sim\RaN{}^{0.5}$, consistent with the scaling in classic \RBC{} \citep{Ahlers2009} and suggesting the buoyancy flow generated by the top-bottom temperature difference determines the scale of flow speed. \tR{We notice that $\ReN{}$, compared to $\NuN{}$, is less sensitive to the change of $\NuN{}_\perp$ at high $\RaN{}$ -- an observation worth further examination.}

The circulation of a perturbed \RBC{} has a non-trivial dependence on $\RaN{}$ and $\NuN{}_\perp$. Shown in \cref{fig3}(c), the time-averaged total angular momentum $|L|$ has a 0.50 power law scaling with $\RaN$ when $\NuN_\perp$ is fixed low, consistent with the scaling of $\ReN{}\sim\RaN{}^{0.5}$, suggesting that the circulation rate is proportional to the flow speed. When $\NuN_\perp$ becomes higher, the scaling deviates and approaches an exponent of 0.34, perhaps due to the development of thin boundary layers near vertical walls seen in \cref{fig2}(b)-(d). This change of scaling for $|L|$ demands a more detailed analysis of the boundary layer structures, which awaits future investigations to justify. On the other hand, $|L|$ is an increasing function of $\NuN{}_\perp$ when holding a constant $\RaN{}$ $<$ $2\times10^5$, but this monotonicity is reversed for $\RaN{}>2\times10^5$ -- increasing horizontal flux $\NuN{}$ instead leads to a decreased total angular momentum. It is known that the large-scale circulation only exists beyond a sufficiently large $\RaN{}$ (around $10^7$), and very little circulation exists for low $\RaN{}$ convection. Therefore, the addition of horizontal flux to low-$\RaN{}$ convection generates an otherwise nonexistent circulation and hence increases the total flow angular momentum. At high $\RaN{}$, increasing $\NuN{}_\perp$ does not change $\ReN{}$ significantly [\cref{fig3}(b)] but the flow becomes confined in a boundary layer [\cref{fig2}(d)], so $|L|$ consequently reduces as the bulk's contribution to the angular momentum integral gradually diminishes.

\subsection{Relaxation of a perturbed \RBC{}}
\label{sec3.3}
\begin{figure}
  \centerline{\includegraphics[width=.9\textwidth]{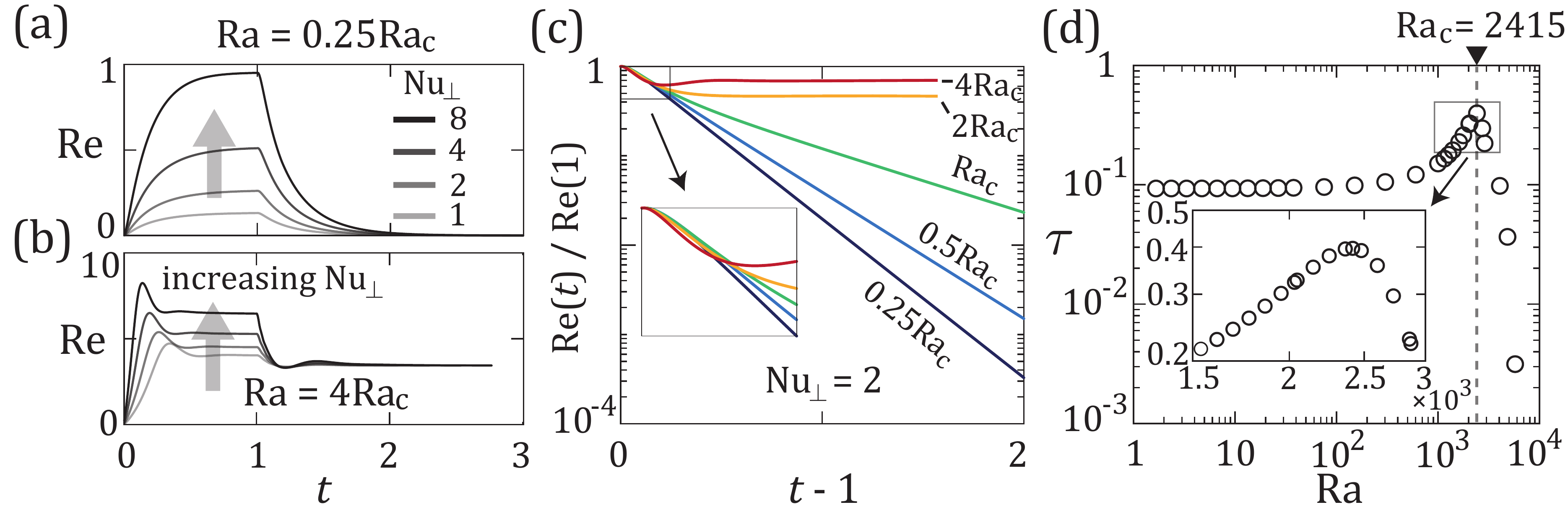}}
  \caption{Perturbing \RBC{} with a transient horizontal flux. An imposed horizontal flux during $t\in(0,1)$ is removed after $t=1$, allowing the system to relax back to the classical Rayleigh Be\'nard configuration. (a) Reynolds number increases in below-onset \RBC{} when a horizontal flux is added, and fluid motion diminishes after this flux is removed. (b) Reynolds number in the above-onset system decreases to the usual level of \RBC{} after the perturbation is removed. (c) The fluid motion decays exponentially in the below-onset \RBC{} after the perturbation is removed. (d) The relaxation time $\tau$ reaches a maximum at $\RaN_{\mbox{c}} = 2415$, approaches a constant as $\RaN{}\to 0$, and decreases rapidly when $\RaN{}>\RaN_{\mbox{c}}$. $\NuN{}_\perp = 2$ in (c)-(d); cases with different values of $\NuN{}_\perp$, shown in (a)-(b), yield similar $\tau$ at a fixed $\RaN$. }
\label{fig4}
\end{figure}

In the following examples, we temporally impose a horizontal heat flux during the time $t\in(0,1)$, and then turn it off at $t=1$ (corresponding to the diffusion time $L^2 /\kappa$) so the system is allowed to return to the configuration of classical \RBC{}. \Cref{fig4}(a) shows how the fluid responds to such perturbations. The Reynolds number for both the below-onset [$\RaN < \RaN_{\mbox{c}}$, \cref{fig4}(a)] and above-onset [$\RaN > \RaN_{\mbox{c}}$, \cref{fig4}(b)] states increases while the horizontal flux is applied. This is expected as there is a horizontal temperature difference in the fluid, which results in convective motion without threshold.

When the horizontal flux is turned off, the flow speed, represented by the Reynolds number $\ReN(t)$, relaxes to that of \RBC{}: In \cref{fig4}(a), the flow velocity drops to 0 as the system is below onset; in \cref{fig4}(b), the flow velocity decreases to a constant $\ReN(\infty)$. \tR{This relaxation is further demonstrated in \cref{fig4}(c). For $\RaN{}<\RaN{}_{\mbox{c}}$, the magnitude of flow speed is found to decay exponentially after removing the horizontal flux, leading to straight lines in the semi-logarithmic plot of \cref{fig4}(c). The slope of these lines shows the rate of decay, and we define a relaxation time $\tau$ that is the negative inverse of the slope. In the case of $\RaN{}>\RaN{}_{\mbox{c}}$, one can also calculate $\tau$ from $\ReN(1) - \ReN(\tau) = e^{-1} [\ReN(1) - \ReN(\infty)]$, providing that $\ReN(\infty)$ can be accurately determined. Shown in \cref{fig4}(d), $\tau$ is a non-monotonic function of $\RaN{}$ that peaks at a critical value determined as $\RaN_{\mbox{c}} = 2415$, where \RBC{} takes the longest time to relax due to a phenomenon known as critical slowing down \citep{RevModPhys.49.435}.} This can also be seen in the \textit{inset} of \cref{fig4}(c), where the $\RaN=\RaN_{\mbox{c}}$ curve (green) decays the slowest. It is worth noting that, at the same $\RaN_{\mbox{c}}$, the Nusselt enhancement $f$ also reaches a maximum when holding $\NuN_\perp$ constant [\textit{inset} of \cref{fig3}(a)]. 

In the below-onset system, the relaxation of temperature and flow fields is closely associated with the dissipation of energy through the diffusion of heat and momentum. In the extreme case of $\RaN\to0$, \tR{the flow speed decreases to 0 as the driving term in \cref{sec-eqn} vanishes}, and \cref{temp-eqn} becomes the heat equation, $\partial\theta/\partial t = \Delta \theta$, whose solution decays exponentially in time. The relaxation time of this exponential decay is set by the initial condition of $\theta$ and geometry, therefore independent of $\RaN$. As the flow field is only driven by the temperature field, the magnitude of flow speed shall exhibit the same exponential decay in time. This confirms the exponential decay in \cref{fig4}(c), and explains why the relaxation time $\tau$ in \cref{fig4}(d) is almost constant for small $\RaN$. Consequently, the dimensional relaxation time $T_\tau = L^2 \tau /\kappa \propto  L^2 /\kappa$ has the same scaling as pure thermal diffusion in the limit of $\RaN\to0$. 

Above $\RaN_{\mbox{c}}$, the relaxation time in \cref{fig4}(d) decreases rapidly with $\RaN{}$, as the vertical temperature gradient sustains fluid motion and the perturbation applied during $t\in(0,1)$ is quickly ``washed away".

\section{Discussion}
\label{sec4}
In this work, we numerically explore the effects of an additional horizontal heat flux in \RBC{}. We found that, in 8 decades of $\RaN{}$, this horizontal heat flux induces fluid motion, modifies flow structures, and increases the Nusselt number. We observe a monotonic response in vertical heat flux when the horizontal heat flux is added, for convection both well-below ($\RaN{}\sim 1$) and well-beyond ($\RaN{}\sim 10^8$) onset, and this allows us to directly control the vertical heat transfer rate of \RBC{} in a diverse range of parameters. By adding a horizontal flux, convection can also be initiated in an otherwise below-onset, conductive state. Once this flux is removed, fluid motion decays exponentially and the system returns to its equilibrium. Right at the critical Ra, the Rayleigh-B\'enard system is most sensitive to the perturbation of the horizontal flux, where the highest Nu amplification ratio $f$ and the longest relaxation time $\tau$ are reached. These nonlinear, non-monotonic behaviors of perturbed \RBC{} could find future applications in the design of more complicated thermal devices. 

\begin{figure}
  \centerline{\includegraphics[width=\textwidth]{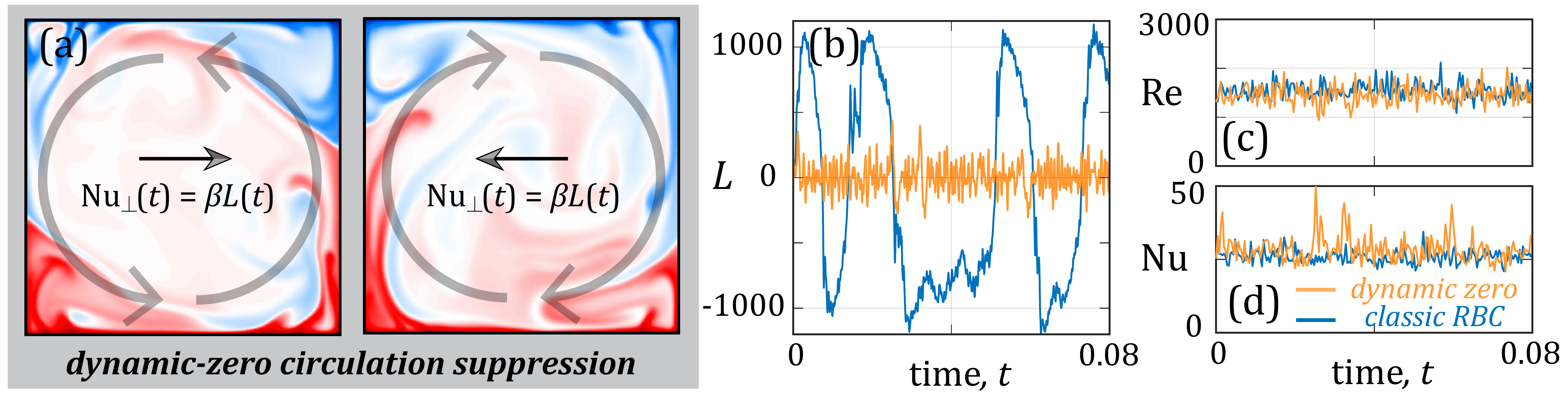}}
  \caption{Actively controlling the \RBC{} with a horizontal flux that depends on the strength of large-scale circulation, $\NuN_\perp = \beta L$. (a) Two moments of the temperature fields and the dynamic-zero control. (b) The total angular momentum $L$ stays close to 0 under the dynamic-zero control regime. (c) The Reynolds number of the controlled \RBC{} is slightly reduced, while (d) the Nusselt number stays unchanged. Classic \RBC{} (blue) data in (b)-(d) is the same as \cref{fig1}. Dynamic-zero regime in (a)-(d) has the same $\RaN = 10^8$ and $\PrN = 4.4$, and the control parameter is \tR{$\beta = 0.128$}. Video of (a) is included in the Supplements.}
\label{fig5}
\end{figure}

As one example of achieving active control, we demonstrate a simple ``dynamic-zero" mechanism to reduce the net large-scale circulation of \RBC{} through adjusting the magnitude and direction of the horizontal flux. Shown in \cref{fig5}(a), we allow the strength of horizontal flux to depend on the total angular momentum such that $\NuN_\perp(t) = \beta L(t)$, where $\beta$ is a positive constant. With this control, a time-dependent, horizontal temperature gradient is imposed so the resulting buoyancy torque counteracts the existing large-scale circulation. Two moments of such active control are shown in \cref{fig5}(a) and its video can be found in the Supplements. Indeed, \cref{fig5}(b) shows that the magnitude of total angular momentum $L$ is suppressed through this dynamic-zero approach, \tR{suggesting a reduced net circulation even though the fluid motion persists and the Reynolds number is only slightly decreased as shown in \cref{fig5}(c). This control mechanism might not be the optimal way of bringing the total angular momentum close to zero, as it does not account for the response time due to thermal inertia. Better control mechanisms, such as the PID controller, can probably suppress the net circulation more efficiently. Nonetheless, our simple mechanism works well as shown in \cref{fig5}(b), considering the total angular momentum is suppressed with a weak control signal. This control signal is given by the orange data in \cref{fig5}(b) multiplied by $\beta = 0.128$. } 

We would naturally ask, is the Nusselt number reduced without the large-scale circulation? It turns out that the Nusselt number in \cref{fig5}(d) stays unchanged, suggesting that the circulation contribution is not significant. However, we have to point out that our dynamic-zero regime actively supplies horizontal flux to the \RBC{}, which is known to increase its $\NuN$. \tR{As we cannot isolate this factor, a more delicate approach is needed for future investigation of large-scale circulations. Experimentally, it is also difficult to simultaneously inject heat through one sidewall while extracting the same amount of heat from the other, so we have instead built an experiment to investigate the influence of sidewall heating (no cooling) in \RBC{} \citep{mac2022controlling}. To control the large-scale circulation there, a modification could include two heating sidewalls, so the direction of large-scale circulation can be controlled by adjusting the heating power on each side. }

Nonetheless, active control is proven possible for the high $\RaN{}$ convection and it can successfully eliminate the otherwise persistent large-scale circulation. Among many avenues for harnessing \RBC{}, the addition of horizontal flux is perhaps the simplest way of regulating one heat flux ($\NuN{}$) with another ($\NuN{}_\perp$), therefore serving a thermal purpose quite similar to the celebrated electronic transistor, and further bridging the thermal and electrical analogies. 

\section*{Acknowledgement}
J.M.H. and J.Z. thank Kaizhe Wang and Zhuang Su for useful discussions. J.M.H. acknowledges support from the ``Chenguang Program" of Shanghai Education Development Foundation and Shanghai Municipal Education Commission (Grant 20CG72). J.Z. acknowledges support from the National Natural Science Foundation of China (Grant NSFC11472106), NYU Shanghai, and partial support by Tamkeen under the NYU Abu Dhabi Research Institute (Grant CG002). 

\section*{Declaration of Interests}
The authors report no conflict of interest.

\section*{Supplementary Materials}
Supplementary movies are available at \url{https://math.nyu.edu/~jinzi/research/HorizontalFlux/Movie}.

\bibliographystyle{tex-library/jfm}
\bibliography{manuscript}

\begin{thebibliography}{31}
\expandafter\ifx\csname natexlab\endcsname\relax\def\natexlab#1{#1}\fi

\bibitem[Ahlers {\em et~al.\/}(2009)Ahlers, Grossmann \& Lohse]{Ahlers2009}
{\sc Ahlers, G., Grossmann, S. \& Lohse, D.} 2009 Heat transfer and large scale
  dynamics in turbulent {R}ayleigh-{B}{\'e}nard convection. {\em Rev. Mod.
  Phys.\/} {\bf 81}~(2), 503.

\bibitem[Bao {\em et~al.\/}(2015)Bao, Chen, Liu, She, Zhang \& Zhou]{Bao2015}
{\sc Bao, Y., Chen, J., Liu, B.-F., She, Z.-S., Zhang, J. \& Zhou, Q.} 2015
  Enhanced heat transport in partitioned thermal convection. {\em J. Fluid
  Mech.\/} {\bf 784}.

\bibitem[Belmonte {\em et~al.\/}(1994)Belmonte, Tilgner \&
  Libchaber]{Belmonte1994}
{\sc Belmonte, A., Tilgner, A. \& Libchaber, A.} 1994 Temperature and velocity
  boundary layers in turbulent convection. {\em Physical Review E\/} {\bf
  50}~(1), 269.

\bibitem[Belmonte {\em et~al.\/}(1995)Belmonte, Tilgner \&
  Libchaber]{Belmonte1995}
{\sc Belmonte, A., Tilgner, A. \& Libchaber, A.} 1995 Turbulence and internal
  waves in side-heated convection. {\em Phys. Rev. E\/} {\bf 51}, 5681--5687.

\bibitem[Busse(1978)]{busse1978non}
{\sc Busse, F.} 1978 Non-linear properties of thermal convection. {\em Reports
  on Progress in Physics\/} {\bf 41}~(12), 1929.

\bibitem[Childress(2009)]{Childress2009}
{\sc Childress, S.} 2009 {\em {An Introduction to Theoretical Fluid
  Mechanics}\/}. {\em Courant Lecture Notes in Mathematics\/} . Courant
  Institute of Mathematical Sciences.

\bibitem[Chong {\em et~al.\/}(2017)Chong, Yang, Huang, Zhong, Stevens,
  Verzicco, Lohse \& Xia]{Chong2017}
{\sc Chong, K.~L., Yang, Y., Huang, S.-D., Zhong, J.-Q., Stevens, R. J. A.~M.,
  Verzicco, R., Lohse, D. \& Xia, K.-Q.} 2017 Confined {Rayleigh}-{B}{\'e}nard,
  rotating {Rayleigh}-{B}{\'e}nard, and double diffusive convection: A unifying
  view on turbulent transport enhancement through coherent structure
  manipulation. {\em Physical Review Letters\/} {\bf 119}~(6), 064501.

\bibitem[Du \& Tong(1998)]{PhysRevLett.81.987}
{\sc Du, Y.-B. \& Tong, P.} 1998 Enhanced heat transport in turbulent
  convection over a rough surface. {\em Phys. Rev. Lett.\/} {\bf 81}, 987--990.

\bibitem[Evans(2010)]{Evans2010}
{\sc Evans, L.~C.} 2010 {\em {Partial Differential Equations}\/}. {\em Graduate
  studies in mathematics\/} . American Mathematical Society.

\bibitem[Grossmann \& Lohse(2000)]{grossmann2000scaling}
{\sc Grossmann, S. \& Lohse, D.} 2000 Scaling in thermal convection: a unifying
  theory. {\em Journal of Fluid Mechanics\/} {\bf 407}, 27--56.

\bibitem[Guo {\em et~al.\/}(2015)Guo, Zhou, Cen, Qu, Lu, Sun \& Shang]{Guo2015}
{\sc Guo, S.-X., Zhou, S.-Q., Cen, X.-R., Qu, L., Lu, Y.-Z., Sun, L. \& Shang,
  X.-D.} 2015 The effect of cell tilting on turbulent thermal convection in a
  rectangular cell. {\em Journal of Fluid Mechanics\/} {\bf 762}, 273--287.

\bibitem[H\'ebert {\em et~al.\/}(2010)H\'ebert, Hufschmid, Scheel \&
  Ahlers]{PhysRevE.81.046318}
{\sc H\'ebert, F. m.~c., Hufschmid, R., Scheel, J. \& Ahlers, G.} 2010 Onset of
  {R}ayleigh-{B}\'enard convection in cylindrical containers. {\em Phys. Rev.
  E\/} {\bf 81}, 046318.

\bibitem[Hohenberg \& Halperin(1977)]{RevModPhys.49.435}
{\sc Hohenberg, P.~C. \& Halperin, B.~I.} 1977 Theory of dynamic critical
  phenomena. {\em Rev. Mod. Phys.\/} {\bf 49}, 435--479.

\bibitem[Hou \& Li(2007)]{Hou2007}
{\sc Hou, T.~Y. \& Li, R.} 2007 Computing nearly singular solutions using
  pseudo-spectral methods. {\em Journal of Computational Physics\/} {\bf
  226}~(1), 379--397.

\bibitem[Howle(1997)]{howle1997active}
{\sc Howle, L.~E.} 1997 Active control of {R}ayleigh--{B}{\'e}nard convection.
  {\em Physics of Fluids\/} {\bf 9}~(7), 1861--1863.

\bibitem[Huang \& Zhang(2022)]{mac2022controlling}
{\sc Huang, J.~M. \& Zhang, J.} 2022 Controlling thermal convection with side
  heating. {\em arXiv preprint arXiv:2103.04042\/} .

\bibitem[Koschmieder(1993)]{koschmieder1993benard}
{\sc Koschmieder, E.~L.} 1993 {\em B{\'e}nard cells and {T}aylor vortices\/}.
  Cambridge University Press.

\bibitem[Nguyen {\em et~al.\/}(2021{\natexlab{{\em a\/}}})Nguyen, Abouezzi \&
  Ristroph]{nguyen2021early}
{\sc Nguyen, Q.~M., Abouezzi, J. \& Ristroph, L.} 2021{\natexlab{{\em a\/}}}
  Early turbulence and pulsatile flows enhance diodicity of {T}esla’s
  macrofluidic valve. {\em Nature Communications\/} {\bf 12}~(1), 1--11.

\bibitem[Nguyen {\em et~al.\/}(2021{\natexlab{{\em b\/}}})Nguyen, Huang,
  Zauderer, Romanelli, Meyer \& Ristroph]{nguyen2021tesla}
{\sc Nguyen, Q.~M., Huang, D., Zauderer, E., Romanelli, G., Meyer, C.~L. \&
  Ristroph, L.} 2021{\natexlab{{\em b\/}}} Tesla's fluidic diode and the
  electronic-hydraulic analogy. {\em American Journal of Physics\/} {\bf
  89}~(4), 393--402.

\bibitem[Niemela {\em et~al.\/}(2000)Niemela, Skrbek, Sreenivasan \&
  Donnelly]{niemela2000turbulent}
{\sc Niemela, J., Skrbek, L., Sreenivasan, K. \& Donnelly, R.} 2000 Turbulent
  convection at very high {R}ayleigh numbers. {\em Nature\/} {\bf 404}~(6780),
  837--840.

\bibitem[Peyret(2002)]{Peyret2002}
{\sc Peyret, R.} 2002 {\em {Spectral Methods for Incompressible Viscous
  Flow}\/}. {\em Applied Mathematical Sciences\/} . Springer New York.

\bibitem[Pursglove(1960)]{Pursglove1960}
{\sc Pursglove, S.~D.} 1960 Fluid ``transistor" circuits. {\em Science and
  Mechanics Magazine\/} {\bf 06}, 81--83.

\bibitem[Schlichting \& Gersten(2016)]{Schlichting2016}
{\sc Schlichting, H. \& Gersten, K.} 2016 {\em {Boundary-Layer Theory}\/}.
  Springer Berlin Heidelberg.

\bibitem[Schmalzl {\em et~al.\/}(2002)Schmalzl, Breuer \&
  Hansen]{schmalzl2002influence}
{\sc Schmalzl, J., Breuer, M. \& Hansen, U.} 2002 The influence of the
  {P}randtl number on the style of vigorous thermal convection. {\em
  Geophysical \& Astrophysical Fluid Dynamics\/} {\bf 96}~(5), 381--403.

\bibitem[Schmalzl {\em et~al.\/}(2004)Schmalzl, Breuer \&
  Hansen]{schmalzl2004validity}
{\sc Schmalzl, J., Breuer, M. \& Hansen, U.} 2004 On the validity of
  two-dimensional numerical approaches to time-dependent thermal convection.
  {\em EPL (Europhysics Letters)\/} {\bf 67}~(3), 390.

\bibitem[Sch{\"o}nfeld(1954)]{schonfeld1954analogy}
{\sc Sch{\"o}nfeld, J.~C.} 1954 Analogy of hydraulic, mechanical, acoustic and
  electric systems. {\em Applied Scientific Research, Section A\/} {\bf 3}~(1),
  417--450.

\bibitem[Stevens {\em et~al.\/}(2009)Stevens, Zhong, Clercx, Ahlers \&
  Lohse]{Stevens2009}
{\sc Stevens, R. J. A.~M., Zhong, J.-Q., Clercx, H. J.~H., Ahlers, G. \& Lohse,
  D.} 2009 Transitions between turbulent states in rotating
  {R}ayleigh-{B}{\'e}nard convection. {\em Phys. Rev. Lett.\/} {\bf 103},
  024503.

\bibitem[Trefethen(2000)]{trefethen2000spectral}
{\sc Trefethen, L.~N.} 2000 {\em Spectral methods in MATLAB\/}. SIAM.

\bibitem[Wang {\em et~al.\/}(2018)Wang, Wan, Yan \&
  Sun]{PhysRevFluids.3.113503}
{\sc Wang, Q., Wan, Z.-H., Yan, R. \& Sun, D.-J.} 2018 Multiple states and heat
  transfer in two-dimensional tilted convection with large aspect ratios. {\em
  Phys. Rev. Fluids\/} {\bf 3}, 113503.

\bibitem[Zhang {\em et~al.\/}(2020)Zhang, Xia, Zhou \&
  Chen]{zhang2020controlling}
{\sc Zhang, S., Xia, Z., Zhou, Q. \& Chen, S.} 2020 Controlling flow reversal
  in two-dimensional {R}ayleigh--{B}{\'e}nard convection. {\em Journal of Fluid
  Mechanics\/} {\bf 891}.

\bibitem[Zhong \& Ahlers(2010)]{Zhong2010a}
{\sc Zhong, J.-Q. \& Ahlers, G.} 2010 Heat transport and the large-scale
  circulation in rotating turbulent {R}ayleigh-{B}{\'e}nard convection. {\em J.
  Fluid Mech.\/} {\bf 665}, 300--333.

\end{thebibliography}

\end{document}